\documentclass{sigchi}

\usepackage{balance}  
\usepackage{graphics} 
\usepackage{times}    
\usepackage{url}      
\usepackage[frenchb]{babel}

\makeatletter
\def\url@leostyle{%
  \@ifundefined{selectfont}{\def\UrlFont{\sf}}{\def\UrlFont{\small\bf\ttfamily}}}
\makeatother
\def\pprw{8.5in}
\def\pprh{11in}

\renewcommand{\thesection}{\Arabic{section}} 

\usepackage[pdftex]{hyperref}
\hypersetup{
pdftitle={SIGCHI Conference Proceedings Format},
pdfauthor={LaTeX},
pdfkeywords={SIGCHI, proceedings, archival format},
bookmarksnumbered,
pdfstartview={FitH},
colorlinks,
citecolor=black,
filecolor=black,
linkcolor=black,
urlcolor=black,
breaklinks=true,
}

\toappear{\scriptsize 

{\emph{IUI 2017, March 13--16, 2017, Limassol, Cyprus.} } \\

http://dx.doi.org/10.1145/3025171.3025226}

\begin{document}

\title{Inline Co-Evolution between Users and Information Presentation for Data Exploration}

\numberofauthors{4}
\author{
  \alignauthor Landy Rajaonarivo \\
  \email{rajaonarivo@enib.fr}\\
  \alignauthor Matthieu Courgeon\\
  \email{courgeon@enib.fr}\\
   \alignauthor Eric Maisel\\
   \email{maisel@enib.fr}\\
   \alignauthor Pierre De Loor\\
   \email{deloor@enib.fr}\\
   \affaddr{Lab-STICC, ENIB, UBL}\\
    \affaddr{29280, Plouzan\'e, FRANCE}\\
}


\maketitle

\begin{abstract}
This paper presents an intelligent user interface model dedicated to the exploration of complex databases. This model is implemented on a 3D metaphor: a virtual museum. In this metaphor, the database elements are embodied as museum objects. The objects are grouped in rooms according to their semantic properties and relationships and the rooms organization forms the museum. Rooms’ organization is not predefined but defined incrementally by taking into account not only the relationships between objects, but also the user’s centers of interest. The latter are evaluated in real-time through user interactions within the virtual museum. This interface allows for a personal reading and favors the discovery of unsuspected links between data. In this paper, we present our model's formalization as well as its application to the context of cultural heritage.


\end{abstract}


\section{Introduction}
This paper is a part of a project aiming at designing tools favoring the sensemaking of a user from his interactions with numerous information stored in a database through a virtual environment. Sensemaking is a concept that comes from the organizational theory \cite{Weick1995} and inspired by the enactive field of cognitive science \cite{Varela-1993}. Enaction considers that mental representations of a living being emerge from his interactions with its environment. 
Moreover, since the living being acts on the environment, it evolves and then, representations result from a circular and dynamical process.

In the case where the living being is a human person and the environment is populated with data embodied with graphical objects (for instance, in a virtual environment), sensemaking involves an exploration with coming and going through the environment rather than a research of some precise data. The discovery of links between data during his exploration constitutes progressively a sense for a user.

However, there are different kinds of links between data as well as different users that will adopt different points of view on these data. In many cases, the number of potential links between data is very important. It is therefore necessary to present to the user only data linked together according to the user's concern. There are as many representations of data organization as there are users. Moreover, these representations can modify user's point of view during his exploration. Consequently, the choice of data presented to the user should be constructed in line as well as the organization of this presentation. To achieve this goal, the tool that allows the exploration of data should evaluate the user interests in real time. It should be able to choose new data to present to the user at any time. The result is a co-evolution between the user point of view and the data presented to him.

From these considerations, we propose a tool that constructs a virtual environment which presents data progressively, according to the real-time evaluation of user's interests. In this paper, it is applied to cultural heritage objects. However, the model is generic and we will precise in which conditions it is possible to apply it to any domains.


This paper is divided into height parts: the first presents the related works, the second describes our approach, the third introduces our data, the fourth to the sixth present the dynamic evolution, the seventh highlights the taking into account of  the  memory maintenance  in our system and the data organization in the museum, and the last parts present our future work and the conclusion.
\section{Related works}

We identified two research domains related to our work: firstly, researches on data visualization aiming at enhancing user's understanding, and secondly, researches on user's models dedicated to adaptive human-machine interfaces.
\subsection{Data visualization}
Visualization is a key element to help a user to analyze data. It allows a global view of the data, organized according to criteria relative to a model elaborated from a point of view. One way to present data is to use a hierarchical structure. For instance, \cite{heer2004doitrees} propose to use a tree constituted by nodes and edges. The nodes represent data categories while the edges represent the hierarchical relationships between the categories. The display depends on the nodes list selected by the user during his navigation. He chooses some kinds of information before obtaining the corresponding representation. \cite{yang2002interring} propose a radial representation, for which a disc area represents the data set, and is divided into many sectors. Each sector is labeled by one description and one color which characterizes one category. The more a category is close the user's interest, the more its corresponding sector is large. User's interest is evaluated in terms of the selections and the zooms which are performed by the user. \\
For \cite{jankun2003moiregraphs}, data are represented by images and are positioned on a spiral. The image selected by the user is placed on the center of the spiral. There is a hierarchy between the data used to position child images on child spirals. The geometrical distance between the central image and each other images corresponds to a hierarchical distance between them. The more the distance of an image from the central image is large, the more its size is small. The user can select any image of interest with the computer mouse to bring a displacement of this image at the center of the screen. Then the position of the other images is re-organized as well as the different spirals.\\
All these approaches use quite abstract representations without a link with the daily environment of an individual. The appropriation of the system by the users is made delicate. For this reason, some approaches use metaphors reproducing a familiar environment, generally through visualization techniques in 3D. For instance, \cite{Card1996} use a virtual library where each book represents one data and each shelf represents one category. Wise et al. present the data on a topographic card \cite{Wise1999}.  Young \cite{Young1997} and Sparacino \cite{Sparacino1999} use another metaphor: that of a city in which the data are represented by buildings, which are grouped into neighborhood. The relations between the data are represented by paths or bridges.\\
Other works as \cite{cho2016vairoma} focused on data categorization according to time, space and events, which allows the user to adapt the representation. Temporal information are presented on a timeline, spatial information on a card, and event information on a disc.
\subsection{User adaptation}
Many works in e-commerce and data exploration focus on finding ways to adapt information representation according to user's center of interest. Almost of these approaches estimate this center of interest from informations consulted during navigation. \cite{paterno2000effective} propose to classify users among three categories: expert, medium, beginner. Each category is associated with a level of information details to be presented to the user, as well as with a specific navigation mode. The system defines and changes the user category by analyzing the information that he consults. \cite{li2007dynamic} present a web site dedicated to scheduling. It allows suggestions according to user's center of interest. These centers of interest are determined from the sites that were previously visited. One similar approach is proposed by Cecilia di Sciascio in \cite{di2016rank}. It consists to not only sort in real time the documents in terms of their relevance compared to the user's centers of interest but also to suggest him some keywords. The user's centers are estimated from the informations consulted during his navigation. \\
Lenzmann et al. \cite{lenzmann1995intelligent} and \cite{DBLP:conf/ijcai/LenzmannW95} present an adaptive virtual interface using a virtual agent which accomplishes some tasks. According to the user's satisfaction level which is evaluated from his gestural and his vocal commands, the agent improves progressively the quality of the service asked.
Finally, other approaches as \cite{mezhoudi2013user} propose adaptive interfaces based on machine learning and user feedback.


\section{Our approach}



\begin{figure}
\includegraphics[width=8.5cm]{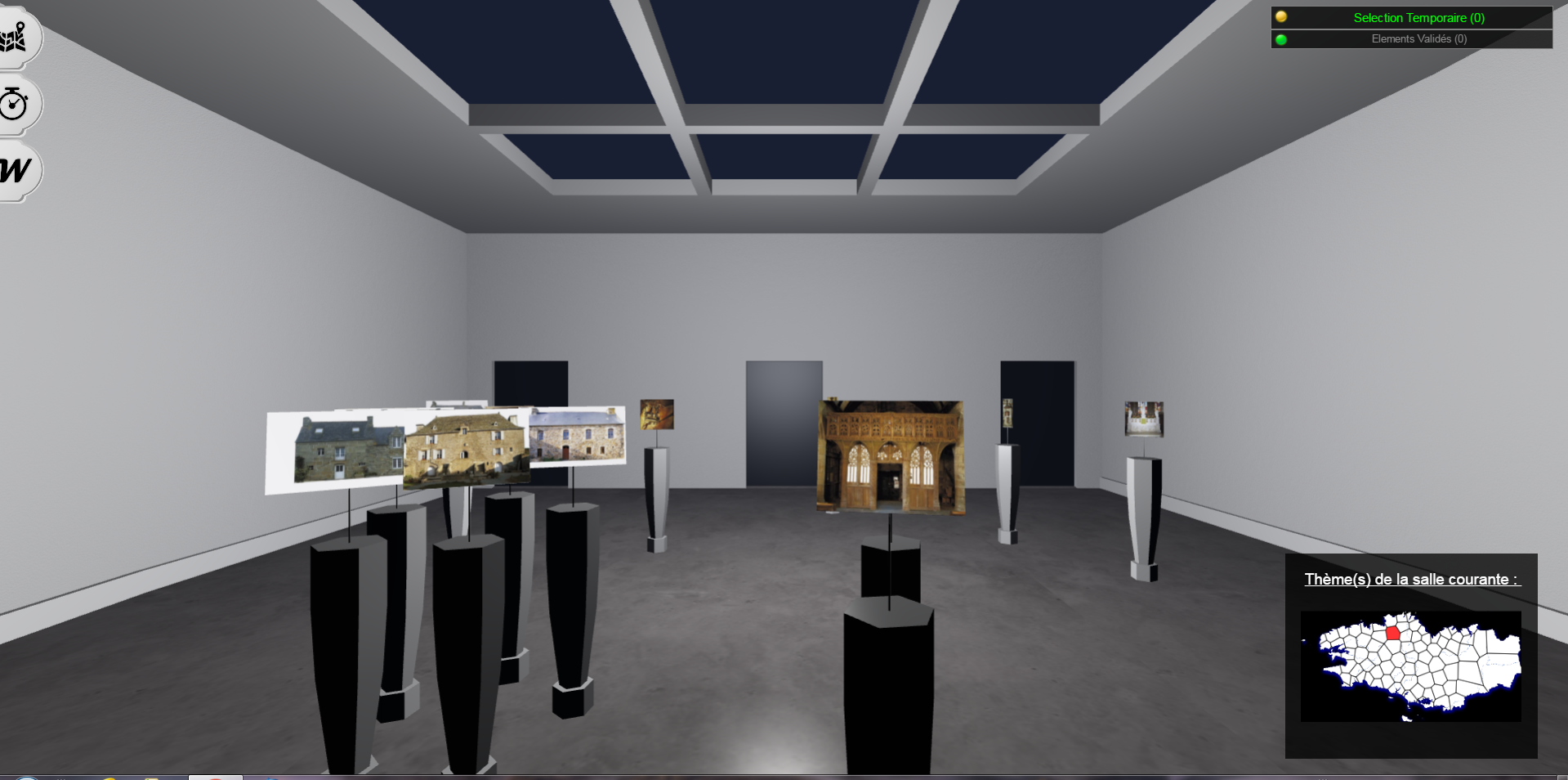}
\caption{Growing virtual museum}
\label{figure:antimoiSnapShot}
\end{figure}

Our data are composed of the heritage objects pictures, text descriptions and keywords which are defined by linguists. We use a museum metaphor to implement the co-evolution principle. Each room presents pictures of some cultural heritage objects. 
Figure \ref{figure:antimoiSnapShot} shows the user's view of the virtual museum.
These objects may have numerous links together. For example, they have been discovered in the same place (topological links)  or they were able to be created at the same time (chronological links). A lot of anthropological links are relative to their usage or signification during certain period of History. 
During a stroll through the museum, new rooms appear, according to the evaluation of user's interest. This new rooms are populated with new objects. The user can guide the system as well as the system guides the user by proposing new rooms and new objects. Some objects are discovered by the user through links he is not aware of. The system is thus capable of improving user's knowledge according to his focus and can introduce serendipity. 

Three interactive tools allow the user to indicate which kind of information he wishes to focus on. These tools also give some feedback to the user about the system's internal evaluation of the user's interests. 
The \textit{Topos} tool shows a colored map. When the user clicks on an area, it increases this area's relevance within the system evaluation process. 
If one area is important to the user, its color turns green and over time progressively degrades to red.
The \textit{Chronos} tool allows to select a period of time and provides the same colored feedback as the \textit{Topos} tools. The \textit{Thema} tool is an interactive word-cloud. The user can click on words to indicate his interests. The words cloud is then self organized accordingly. A complete explanation of this self organization is provided in \cite{rajaonarivo2016}. Semantic links between cultural objects and words and between words (together) exist and are provided through an ontology dedicated to anthropological reading of cultural heritage. As this ontology is developed by linguist partners of the project, it is a specific work in progress that we will not detail in this paper. For this particular work, we will only consider it as a graph which links concepts and keywords together.  

\begin{figure}
\includegraphics[width=8.5cm]{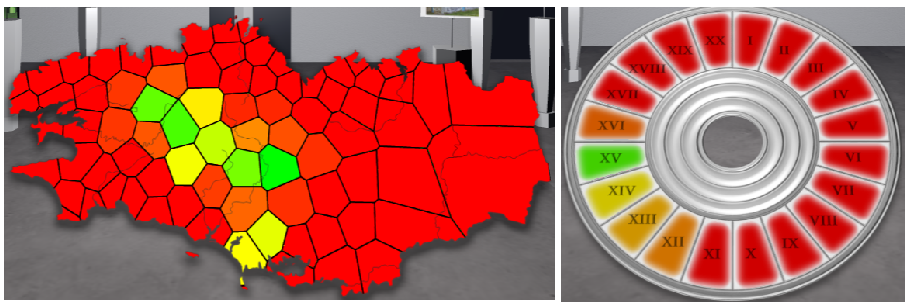}
\caption{On the left : the \textit{topos} tool and on the right the \textit{chronos} tool}
\label{figure:toposChronosTools}
\end{figure}

\section{Data space}
The underlying model of our proposition relies on the following elements:

\begin{itemize}
\item Objects:\\
\og Objects\fg{} are the heritage objects from the database of our industrial partner. Each of them is associated with different elements as an image, a descriptive text, as well as dimensional entities (see below). 


\item Dimension:\\
The dimension are the \textit{time}, the \textit{space}, and the \textit{topic}, and correspond to the categories which are manipulable by the \textit{Chronos}, \textit{Topos}, and \textit{Thema} tools, presented above.
\item Dimensional entities:\\
Each object is characterized by at least three dimensional entities which belong to the three dimensions: \textit{Chronos}, \textit{Topos} and \textit{Thema}. The dimensional entities in the dimension \textit{Chronos} define the period during which the object was created or used, \textit{Topos} dimensional entities correspond to the place where the object was created or used, while dimensional entities of the dimension \textit{Thema} is derived from the ontology and are concepts related to the object. This number of three dimensional entities by object is a minimum. For example, there may be several dimensional entities in the \textit{Chronos} dimension which correspond to important dates for the object (destruction, renovation ...).
\item Distance:\\
Within each dimension, we have defined methods to calculate a distance between two dimensional entities. The temporal distance is defined by the smaller number of century which separates the two dimensional entities of the \textit{Chronos} dimension. The geographical distance between two \textit{Topos} dimensional entities is defined by the length (Territory count) of the shortest path that separate them. Finally, the semantic distance is determined using the ontology. A distance of 1 correspond to a direct semantic link between two concepts of the ontology.
\end{itemize}

\begin{figure}[htb]
\begin{center}
\includegraphics[width=7.5cm]{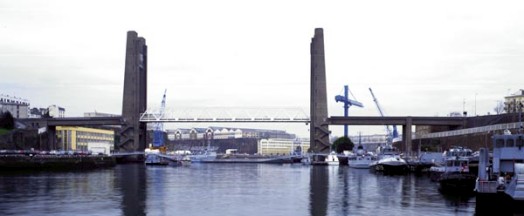}
\caption{Object characterized by three Dimensional entities: $Topos$: Brest, $Chronos$: $XX^{th}$ century, $Thema$: Bridge}
\label{figure:exempleObjet}
\end{center}
\end{figure}
Let:
\begin{itemize}
\item \textit{Ob}=\{$ob_i, i=1..N_{ob}$\}: the set of heritage objects present in the environment where $N_{ob}$ is the size of this set\\
\item \textit{D}=\{$d_i, i=1..N_d$\}: the dimensions set where $N_d$ is the size of this set. In our case, we have three dimensions(\textit{Chronos}, \textit{Topos}, and \textit{Thema}).
\item $DE_{d_j} = \{de_{i},Dim(de_{i}) = d_j \}$: the set of dimensional entities in the dimension $d_j$. $Dim(de)$ is the function which return the dimension of one dimensional entity $ed$. 
\end{itemize}
All these definitions are not domain dependent and are available once the following conditions have been met: having a database associated with ontologies that are structured in term of dimensional entities. Each dimensional entity belong to a dimension.

\section{Virtual Museum Self-dynamic }
\subsection{Museum tree structure}
The virtual museum is composed of a set of rooms which are dedicated to at least one topic. Topics are dimensional entities (from the three dimension set). The rooms are organized according to a tree structure. Each room have one \og parent\fg room, from which the user has entered. Each room has three doors (closed in the beginning) leading to its \og children\fg{} rooms, plus one open door leading to its \og parent\fg room. These three closed doors will open dynamically upon dynamic creation of their respective child-rooms. Each room thus have at most three "children" rooms. To avoid spatial collision of rooms on big museum structure, the rendering engine uses a non euclidean space mechanism that prevents rooms to be rendered simultaneously at the same location.
\subsection{Estimation of user's center of interest}
The estimation of user's center of interest is based on his interactions with the environment during his navigation. We defined 6 types of interactions:
\begin{enumerate}
\item to stand in front of an object
\item to consult text description of an object
\item to put an object in a virtual basket (this basket is used to keep within easy reach some objects that interest the user)
\item to remove an object from the virtual basket
\item to enter in a room
\item to use the tools
\end{enumerate}
Let:
\begin{itemize}
\item $I = \{type_I,DE_I,W_I\}$ an interaction of $type_I$ with a weight $W_I$ on an object or tools concerned by the dimensional entities set $DE_I$ 
\item $t_I$: the time when the interaction $I$ is realized
\item $Trace=\{(I_1,t_{I_1}), (I_2,t_{t_2}), ... (I_n,t_{I_n})\}$ the temporal recording of the interactions that the user realizes during his exploration.
\item $w(I,t)$: the weight associated with the interaction $I$ at time $t$ after it was performed. The temporal evolution of this weight is computed as follows: \\ $w(I, t) \leftarrow W_I * \exp(-\lambda*(t-t_I))$. With this formula, the value of this weight decreases in half after $\frac{ln(2)}{\lambda}$ seconds.
\item $R(de)$ the real-time relevance of the dimensional entity $de$ for the user
\end{itemize}

Each second, the simulation is updated according to the following algorithm:\\
$\forall (I_i,t_{I_i}) \in Trace,$ and $\forall \ de \in DE_{I_i}$ : \\
$R(de) \leftarrow R(de) + w(I_i,t_{I_i})* (R_{de\_max} -R(de))$\\ where $R_{de\_max}$ is a maximal relevance value defined for $de$. \\





Then, to favor serendipity and to help the user to find new objects, we decrease and propagate the relevance of dimensional entities to their neighbors (in their respective dimensions \textit{Topos}, \textit{Chronos} and \textit{Thema}). For that, if the relevance of one dimensional entity is superior than a given threshold, this latter diffuses one part of its relevance to its direct neighbors. To avoid the cyclic diffusion, we prevent one dimensional entity to diffuse its relevance more than once for a certain time: 
\begin{itemize} 
\item Decreasing:\\
$\forall (I,t_I) \in Trace,$ and $\forall de \in DE_I,$ \\
$R(de) = \tau * R_{de\_min} + (1-\tau) * R(de)$ where $\tau$ is the reduction rate and $P_{ed\_min}$ the minimal relevance of $de$.
\item Propagation: \\
Let:
\begin{itemize}
\item $s_d$: the threshold of the relevance diffusion
\item $\gamma$: the rate of the diffusion
\item $de$: a dimensional entity
\item $Ng(de)$: the set of neighbors of $de$. This neighboring is defined by a threshold distance.
\end{itemize}
If $R(de) > s_d, then:$\\
$\forall de_i \in Ng(de), Ng(de_i) = R(de_i) + \frac{\gamma*R(de)}{|V(de)|}$\\
$R(de) \leftarrow (1 - \gamma)*R(de)$\\
This formula is inspired by energy conservation. The relevance lost by one dimensional entity is equal to the sum of that it has diffused to its neighbors.
\end{itemize}
\section{The evolution of the museum structure}
From the users' center of interest, the system self-determines the moments at which a new room should be present to the user: 1) when one $de$ reaches a high pertinence threshold and it is not present in the adjoining rooms or 2) when the user approaches a closed door. When one room is created, we associated to it one topic which is formed by a set of one or two \og Dimensional entities \fg{}.
\section{Objects self-organization}
The objects presented in the room are determined semi-automatically. The system groups a set $G1$ of objects which correspond to the junction of the $de$ associated to the room. A second object group $G2$ is created and correspond to a set of objects having similar topic(distance 1 from the $de$ of room topic), and the group $G3$ formed by the $de$ to distance 2. The room content is formed by the selection of 12 objects belong them 40\% of $G1$, 40\% of $G2$, and 20\% of $G3$. It allows the user discovering thematic related objects.\\
Once these objects are selected, their position in the room is self-organized by distance. All the objects are randomly positioned in the room and there is mass-spring system allowing to position the objects based on their similarity distances. The spring rest length between two objects is proportional to the sum of the distance between their $de$. Once the system stabilized, the position of the objects are normalized in order to enter all objects in the room while respecting the distance between them.
\section{Memory maintenance}
In order to not disorient the user, no room are destructed during the simulation. This allows the user to have references when he wants to turn back onto his path. We note that we could have many rooms having the same themes (and objects). For this, the user can easily find back an object without the need to come back. Obviously this object will appears only if it belongs to the user's center of interest (or if it is chosen by the algorithms randomly to favor serendipity). 

\section{Conclusion and perspectives}

We have presented an approach that allows for the co-evolution between a user knowledge on one domain and the presentation of the informations to him while preserving serendipity. The work presented in this paper illustrates this problem through a particular case: the reading of the heritage for the study of human societies. This reading is an intellectual construction that we have embodied through the mediation of virtual reality, through the adaptive creation of a virtual world, and according to a museum metaphor. The subjective character of this construction is took into account by the adoption of the enaction paradigm which is translated here by the sensori-motor coupling linking the user to the information system. This paradigm is implemented by a process which describe the relevance diffusion of heritage objects by taking into account both the conceptual description of heritage and the user's intentions (inferred from user's interaction within the virtual environment). \\
The implemented system leads to the creation of a virtual environment usable as interface between the user and a heritage database. This interface presents a 3D representation of the information contained in the database as well as the meta-information usable to guide the user in the environment.\\
We intend to continue this work in different directions: firstly, we will develop an evaluation methodology adapted to the developed approach. Several axis must be validated, such as the precision of users' center of interest estimations, navigation, understanding of museum structure, and so on. Then, we will take into account of the user's body. The goal is to give sense to the information according their position compared to the user's body and rather than only using the 3D environment. Finally, we will apply our proposition in the context of social networks, publications of scientific papers,

\bibliographystyle{acm-sigchi}
\bibliography{sample}
\end{document}